\documentclass[a4paper,authoryear,review]{elsarticle}
\pdfoutput=1 
\journal{Journal of Derivatives and Quantitative Studies}

\usepackage{amsmath,amssymb,graphicx,hyperref,natbib,setspace,algorithm}
\usepackage[noend]{algpseudocode}
\usepackage[left=1in,right=1in,top=1.4in,bottom=1.2in]{geometry}

\newcommand{\qtext}[2][\quad]{#1\text{#2}#1}
\newcommand{\vect}[1]{\boldsymbol{#1}}
\newcommand{\mat}[1]{\boldsymbol{#1}}
\newcommand{\Cov}{\boldsymbol{C}}
\newcommand{\Corr}{\boldsymbol{R}}
\newcommand{\ww}{\boldsymbol{w}}

\hypersetup{
	pdfauthor={Jaehyuk Choi and Rong Chen},
	pdfkeywords={risk parity, equal risk contribution, cyclical coordinate descent, Newton method},
	colorlinks=true,
	linkcolor=red,
	citecolor=blue,
	urlcolor=blue,
	bookmarksnumbered=true,
	pdfstartview=
}

\begin{document}
\begin{frontmatter}
\title{Improved iterative methods for solving risk parity portfolio \tnoteref{t1}}
\tnotetext[t1]{\underline{Please cite as:} Choi, J., \& Chen, R. (2022). Improved iterative methods for solving risk parity portfolio. \textit{Journal of Derivatives and Quantitative Studies}, 30(2). \url{https://doi.org/10.1108/JDQS-12-2021-0031}\vspace{1ex}\\
The authors accepted manuscript is deposited under a Creative Commons Attribution Non-commercial 4.0 International (CC BY-NC) licence. This means that anyone may distribute, adapt, and build upon the work for non-commercial purposes, subject to full attribution. If you wish to use this manuscript for commercial purposes, please contact \url{permissions@emerald.com}.}

\date{February 9, 2022}

\author{Jaehyuk Choi\corref{corrauthor}\fnref{ack}}
\ead{jaehyuk@phbs.pku.edu.cn}

\author{Rong Chen}
\ead{chenr18@pku.edu.cn}

\cortext[corrauthor]{Corresponding author \textit{Tel:} +86-755-2603-0568, \textit{Address:} Peking University HSBC Business School, University Town, Nanshan, Shenzhen 518055, China}

\fntext[ack]{The authors would like to thank Sol Kim and Eun Jung Lee (editors), two anonymous referees, and Minsuk Kwak (discussant at the 2021 Korea Derivatives Association Annual Conference) for their valuable comments. Jaehyuk Choi thank Junyoep Park for testing the new methods of this study. This research was supported by the 2021 Korea Derivatives Association and FnGuide research grant. The Python code used in this study can be found at \url{https://github.com/PyFE/PyfengForPapers} and \url{https://github.com/PyFE/PyFENG}.}

\address{Peking University HSBC Business School, Shenzhen, China}

\begin{abstract}
Risk parity, also known as equal risk contribution, has recently gained increasing attention as a portfolio allocation method. However, solving portfolio weights must resort to numerical methods as the analytic solution is not available. This study improves two existing iterative methods: the cyclical coordinate descent (CCD) and Newton methods. We enhance the CCD method by simplifying the formulation using a correlation matrix and imposing an additional rescaling step. We also suggest an improved initial guess inspired by the CCD method for the Newton method. Numerical experiments show that the improved CCD method performs the best and is approximately three times faster than the original CCD method, saving more than 40\% of the iterations.
\end{abstract}
\end{frontmatter}

\section{Introduction} \label{s:intro} \noindent
Optimal portfolio selection is an important question in academia and the financial industry. While there are several traditional allocation methods, such as mean-variance, minimum variance, and equally weighted ($1/N$) portfolios, the risk parity (or equal risk contribution) model has recently gained popularity in the asset management industry. Under the risk parity model, the portfolio weights are selected in such a way as to equalize the contribution of each asset to the portfolio volatility. Like the $1/N$ portfolio, risk parity aims to diversify and overcome the concentration or sensitivity issues found in the mean-variance or minimum variance portfolios. However, risk parity is the $1/N$ portfolio in terms of risk allocation, rather than capital allocation. 

While it is unclear who started using the risk parity strategy in the financial industry, \citet{maillard2010properties} and \citet{qian2011risk} are widely cited as references that have introduced this strategy to academia. There is a growing body of literature on various aspects of the risk parity allocation method. For example, see \citet{chaves2011risk} and \citet{clarke2013risk} for a comparison of risk parity with other asset allocation methods. \citet{kim2020study} study the risk parity model where the covariance is estimated from the XGBoost algorithm. \citet{kim2021reduction} studies the risk parity under the covariance estimation error. The actual performance of funds implementing this strategy is controversial. See \citet{corkery2013fashionable} for more details.

Among the academic research topics related to the risk parity model, this study is primarily concerned with numerical methods to solve risk parity portfolio weights. Solving the portfolio weight for a given return covariance matrix is challenging because an analytical solution is not available. One must resort to numerical methods, and several methods are currently available. \citet{maillard2010properties} formulate the risk parity problem using a sequential quadratic programming (SQP) method. \citet{chaves2012eff} uses the Newton method to solve the multidimensional root, taking advantage of the analytical Jacobian matrix. Although the original Newton method cannot guarantee the weights to be positive, \citet{spinu2013algorithm} resolve the issue by introducing damped iteration steps. A competing method is the cyclical coordinate descent (CCD) algorithm~\citep{griveau2013fast}. Because the CCD method uses quadratic iteration steps, it does not rely on a Jacobian matrix. \citet{bai2016least} propose the alternating linearization methods (ALMs) for solving the risk parity weight in a generalized setting.

According to the literature, the performances of the CCD and Newton methods are comparable. \citet{griveau2013fast} claim that the CCD method outperforms the Newton method when the number of assets is larger than 250. \citet{bouzida2014robust} reports that, while the CCD method is faster than \citet{spinu2013algorithm}'s Newton method, it lacks the robustness for a pool of assets.

This study reviews and improves two algorithms for solving the risk parity portfolio allocation: the CCD and Newton methods. We improve the CCD method by simplifying it to a numerically efficient form. Furthermore, we improve the Newton method by suggesting a new initial guess. Numerical experiments with randomly generated covariance matrices show that our improved CCD method outperforms the other methods in terms of speed and stability for a wide range of portfolio sizes. 

The remainder of this paper is organized as follows. Section~\ref{s:model} introduces the risk parity portfolio and its properties. Section~\ref{s:method} presents the improved root-finding method, and Section~\ref{s:numeric} demonstrates the computational gain of the new methods using numerical experiments. Finally, Section~\ref{s:con} concludes this paper.

\section{Risk parity portfolio}\label{s:model}
\subsection{Notations and conventions} \noindent
We define several notations and operations regarding vectors and matrices to be used in the remainder of this study.
\begin{itemize}
	\item For vector $\vect{x}$ (in boldface), the $i$-th element is denoted by $x_i$ or $(\vect{x})_j$.
	\item For matrix $\mat{A}$ (in boldface), the $(i,j)$ element is denoted by $A_{ij}$ or $(\mat{A})_{ij}$.
	\item Vectors are assumed to be column vectors unless otherwise specified.
	\item $\vect{1}_N$ is the $N\times 1$ column vector filled with $1$'s. $\mat{I}_N$ is an $N\times N$ identity matrix.
	\item The operations, $*$ and $/$, between vectors $\vect{x}$ and $\vect{y}$ are defined to be the element-wise multiplication and division, respectively:
	$$ \vect{x}*\vect{y} = (x_1y_1, \cdots, x_Ny_N)^T \qtext{and} \vect{y}/\vect{x} = (y_1/x_1, \cdots, y_N/x_N)^T
	$$
	\item The operation $*$ between a (column) vector $\vect{x}$ and matrix $\mat{A}$ is defined as
	$$ \vect{x}*\mat{A} = \mat{A}*\vect{x}= \begin{pmatrix}
		x_1 A_{11} & \cdots & x_1 A_{1N} \\
		\vdots & \ddots & \vdots & \\
		x_N A_{N1} & \cdots & x_N A_{NN}
	\end{pmatrix}
	$$
\end{itemize}

\subsection{Condition for risk parity portfolio} \noindent
Let $\vect{\sigma}$ and $\mat{C}$ be the standard deviation vector and covariance matrix, respectively, of the return on $N$ assets in a unit time period. The covariance $\mat{C}$ is symmetric and positive semi-definite, and its diagonal elements are related to $\vect{\sigma}$ by $C_{ii} = \sigma_i^2$. The portfolio of $N$ assets invested with weight $\ww$ has return volatility:
$$ V(\ww) = \sqrt{\ww^T\Cov\ww}.
$$
From Euler's homogeneous function theorem, volatility $V(\ww)$ can be decomposed into the sum of the contributions of each asset as
$$ V(\ww) 
= \sum_i v_i(\ww) \qtext{where}
v_i(\ww) = w_i \frac{\partial V(\ww)}{\partial w_i} = \frac{w_i(\Cov\ww)_i}{V(\ww)}.
$$

Let $b_i$, such that $\sum_i b_i = 1$ and $b_i>0$, be the relative contributions of the $i$-th asset to portfolio volatility. Then, we aim to determine the weight $\ww$ that satisfies
$$ v_i(\ww) = V(\ww)\, b_i \quad\text{for all $i$}.$$
The risk parity portfolio is a special case of the problem above, where the risk contributions are equally divided among the assets: 
$$b_i=1/N \quad\text{for all $i$}.$$ 
Although we deal exclusively with the risk parity case, we use $b_i$ to avoid losing generality. Therefore, the risk parity portfolio weight must satisfy the condition
\begin{equation} \label{e:cond_cov}
	w_i (\Cov \ww)_i = V^2(\ww)\,b_i = (\ww^T\Cov\ww)\,b_i
	\qtext{subject to} w_i\ge 0.
\end{equation}
Here, we impose $w_i\ge 0$ because we are concerned with a long-only portfolio. We refer to \citet{bai2016least} for the risk parity with an unconstrained portfolio.

The solution is not unique because of the homogeneous property of the condition. If $\ww$ is a solution, then $\mu \ww$ for $\mu>0$ is also a solution. The degrees of freedom can be fixed by setting $\sum_k w_k = 1$. The normalized weight $\bar{\vect{w}}$ is obtained by
$$ \bar{\vect{w}} = \frac{\ww}{\lambda} \qtext{for} \lambda = \sum_k w_k.
$$

\subsection{Risk parity condition with correlation matrix} \label{ss:reduced} \noindent
The condition for the risk parity portfolio can be equivalently stated in terms of the correlation matrix~\citep{spinu2013algorithm}. Let $\mat{R}$ be the correlation matrix whose element is computed from the covariance matrix $\mat{C}$:
\begin{equation} \label{e:cor}
	R_{ij} = \frac{C_{ij}}{\sigma_i \sigma_j} = \frac{C_{ij}}{\sqrt{C_{ii} C_{jj}}} \quad \left(R_{ii} = 1\right).
\end{equation} 
Then, the risk parity condition with the correlation matrix is given by
\begin{equation} \label{e:cond_cor}
	w_i (\Corr \ww)_i = (\ww^T\Corr\ww)\,b_i \qtext{subject to} w_i\ge 0.
\end{equation}
If $\ww$ is the solution to the correlation condition in Eq.~\eqref{e:cond_cor}, then $\ww/\vect{\sigma}$ is the solution to the covariance condition in Eq.~\eqref{e:cond_cov}.

Moreover, \citet{spinu2013algorithm} removes the degree of freedom in $\ww$ by taking advantage of the fact that $\ww^T\Corr\ww$ is scalar, although the value is unknown. Without loss of generality, by setting
\begin{equation} \label{e:scale}
	\ww^T \Corr \ww = 1,
\end{equation}
we arrive at the simple risk parity condition:
\begin{equation} \label{e:cond_1}
	w_i (\Corr \ww)_i = b_i \qtext{subject to} w_i\ge 0.
\end{equation}
Once we find the unique root $\vect{w}$ in Eq.~\eqref{e:cond_1}, we can obtain the normalized portfolio weight $\bar{\vect{w}}$ by
$$ \bar{w}_i = \frac{w_i}{\lambda \sigma_i} \qtext{for} \lambda = \sum_k \frac{w_k}{\sigma_k}.
$$
Note that Eq.~\eqref{e:scale} is consistent with Eq.~\eqref{e:cond_1} because it can be obtained by summing Eq.~\eqref{e:cond_1} for all $i$,
\begin{equation}
	\ww^T \Corr \ww = \sum_i w_i (\Corr \ww)_i = \sum_i b_i = 1.
\end{equation}
We will use both Eqs.~\eqref{e:scale} and \eqref{e:cond_1} in Section~\ref{ss:ccd} to improve the original CCD method.

\subsection{A special case solution and initial guess for iterative methods} \noindent
A general solution to the risk parity portfolio is not analytically available. However, an analytical solution exists for a special case in the correlation matrix. When the correlation matrix has the same row sum,
$$\textstyle \sum_j R_{ij} = r \quad\text{for all $i$ and some constant $r$},$$
the constant vector, $\ww = \vect{1}_N$, satisfies Eq.~\eqref{e:cond_cor}, where $b_i = 1/N$.
Note that a special case is typically achieved when the assets have a uniform correlation, that is, $R_{ij} = \rho$ for $i\neq j$ and some $\rho$. This is also the case for $N=2$; thus, $\ww = \vect{1}_N$ is the general solution for allocating the two assets. In terms of the simple condition, Eq.~\eqref{e:cond_1}, $w_i = 1/\sqrt{N r}$ is the corresponding solution for the special case. The actual portfolio weights satisfying the original condition in Eq.~\eqref{e:cond_cov} are given by weights that are inversely proportional to the standard deviation:
\begin{equation} \label{e:init1}
\bar{w}_i = \frac{1/\sigma_i}{\sum_k 1/\sigma_k}.
\end{equation}

Even in general cases where the row sums of $\Corr$ are different, the special case solution serves as a first-order approximation. \citet{chaves2012eff} and \citet{griveau2013fast} use Eq.~\eqref{e:init1} as the initial guess for the iterative methods. \citet[Theorem 3.4]{spinu2013algorithm} further improves the initial guess for the simple condition, $w_i = 1/\sqrt{N r}$, to 
\begin{equation} \label{e:cond_ccd}
\textstyle w_i = 1/\sqrt{\sum_{j,k} R_{jk}} \;\;\text{for all $i$}.
\end{equation}
Note that $\sum_{j,k} R_{jk} = \vect{1}_N^T\, \Corr\, \vect{1}_N > 0$ if $\Corr$ is positive definite.

\section{Methods for solving risk parity} \label{s:method} \noindent
This section reviews and improves the CCD and Newton methods. As stated in Section~\ref{s:intro}, there exist other methods, such as the SQP algorithm~\citep{maillard2010properties} and ALM~\citep{bai2016least}. Although such methods may handle risk parity in a more generalized setting, they have been reported to be slower than the CCD and Newton methods in handling the standard long-only risk parity model. Therefore, we focus on these two methods.

\subsection{The improved CCD method} \label{ss:ccd} \noindent
The original CCD method~\citep{griveau2013fast} aims to solve 
\begin{equation} \label{e:cond_v}
w_i (\Cov \ww)_i = \sqrt{\ww^T\Cov\ww}\,b_i = V(\ww)\,b_i \quad\text{for all $i$}.
\end{equation}
Note that this condition is different from the covariance condition in Eq.~\eqref{e:cond_cov} as $V^2(\ww)$ is replaced with $V(\ww)$. Although the intention is not explicitly stated in the reference, the purpose of the replacement seems to break the homogeneous property of $\ww$; if $\ww$ is a root, $\mu \ww$ for $\mu>0$ is no longer a solution except for $\mu =1$. Therefore, CCD iterations eventually lead to a unique root.

The equation for the $i$-th component can be written in the quadratic form of $w_i$:
$$\textstyle
C_{ii} w_i^2 + \left(\sum_{j\neq i} C_{ij}w_j \right) w_i - V(\ww)\,b_i = 0,
$$
and we use the root formula as an iteration step for $w_i$~\citep[Eq.~(4)]{griveau2013fast} 
\begin{equation} \label{e:ccd}
	w_i \leftarrow \frac{\sqrt{a_i^2 + C_{ii}V(\ww)\,b_i} - a_i}{C_{ii}} \qtext{for} a_i = \dfrac12 \sum_{j\neq i} C_{ij}w_j = \frac{(\Cov \ww)_i - C_{ii}w_i}{2}.
\end{equation}
We select the positive one among the two roots of the quadratic equation to ensure that $w_i\ge 0$. In the CCD method, one iteration involves cyclically updating $w_i$ for $i=1,\cdots, N$. Updating $w_i$, therefore, uses $w_j$ for $1\le j<i$, which was previously updated in the same iteration step. Therefore, the CCD method is known to be more effective than \textit{batch} coordinate descent. \citet{griveau2013fast} starts the iterations with the initial guess in Eq.~\eqref{e:init1}.

We improve the original CCD method in two ways. First, we formulate the CCD with the simple condition in Eq.~\eqref{e:cond_1}. The equation for the $i$-th component in Eq.~\eqref{e:cond_1} can be written in the simpler quadratic form of $w_i$:
$$\textstyle
w_i^2 + \left(\sum_{j\neq i} R_{ij}w_j \right)w_i - b_i = 0.$$
The corresponding iteration step is simplified to
\begin{equation} \label{e:ccd2}
	w_i \;\leftarrow\; \sqrt{a_i^2 + b_i} - a_i \qtext{for} 
	a_i = \dfrac12 \sum_{j\neq i} R_{ij}w_j = \frac{(\Corr \ww)_i - w_i}{2}.
\end{equation}
This new iteration has two advantages because $V(\ww)$ disappears in the new CCD method. One obvious advantage is the reduced computation time for $V(\ww)$. The calculation can be time consuming in the original CCD method because $V(\ww)$ requires $O(N^2)$ operations and must be updated when $w_i$ is updated. The other advantage is not obvious but is more important. In the original CCD, the new $w_i$ on the left-hand side depends on the old $w_i$ through $V(\ww)$ on the right-hand side of Eq.~\eqref{e:ccd}. The updated $w_i$ is not the true root of Eq.~\eqref{e:ccd}. However, in the new CCD method, the new $w_i$ is the exact root of Eq.~\eqref{e:ccd2} because the old $w_i$ does not appear on the right-hand side. Therefore, we expect the convergence to be faster in the improved CCD iteration.

Second, we rescale $\ww$ by
\begin{equation} \label{e:ccd2-rescale}
	\ww \;\leftarrow\; \frac{\ww}{\sqrt{\ww^T \Corr \ww}}.
\end{equation}
at the end of each iteration to ensure Eq.~\eqref{e:scale}. This rescaling step is expected to make the convergence faster by adjusting $\ww$ on average. The new CCD method uses a generalized initial guess, Eq.~\eqref{e:cond_ccd}, from \citet{spinu2013algorithm}. In fact, this can be understood as the result of rescaling from an equal weight $\ww = \vect{1}_N$. 

Finally, we summarize the improved CCD algorithm in Algorithm~\ref{a:ccd}. 
\begin{algorithm}
	\caption{The improved CCD algorithm \label{a:ccd}}
	\begin{algorithmic}
		\State \textbf{Input}: covariance matrix $\Cov$ and error tolerance $\varepsilon$
		\State Calculate $\Corr$ and $\vect{\sigma}$
		\State Initialize $\ww \gets \vect{1}_N/\sqrt{\sum_{j,k} R_{jk}} $
		\While{$\max_i \left| w_i (\Corr \ww)_i - b_i \right| > \varepsilon$}
		\For{$i \leftarrow 1,\cdots,N$}
		\State	$w_i \gets \sqrt{a_i^2 + b_i} - a_i \qtext{for} a_i = \dfrac12 \sum_{j\neq i}R_{ij}w_j$
		\EndFor		
		\State $\ww \gets \ww/\sqrt{\ww^T \Corr \ww}$
		\EndWhile
		\Return $(\ww/\vect{\sigma})/(\sum_k w_k/\sigma_k)$
	\end{algorithmic}
\end{algorithm}

\subsection{Newton method with an improved initial guess} \noindent
\citet{chaves2012eff} and \citet{spinu2013algorithm} use the multidimensional Newton method to find the risk parity portfolio. From Eq.~\eqref{e:cond_1}, the objective function is set as
\begin{gather*}
	\vect{F}(\ww) = \Corr\ww - \frac{\vect{b}}{\ww} \qtext{or}
	F_i(\ww) = \sum_j R_{ij} w_j - \frac{b_i}{w_i}.
\end{gather*}
Then, the root of $\vect{F}(\ww) = 0$ is the risk parity weight. The Jacobian of $\vect{F}(\ww)$ is readily available as follows:
\begin{gather*}
	\nabla \vect{F}(\ww) = \Corr + \mat{I}_N*\frac{\vect{b}}{\ww^2}
	\qtext{or}
	\frac{\partial F_i(\ww)}{\partial w_j} = R_{ij} + \delta_{ij}\frac{b_i}{w_i^2},
\end{gather*}
where $\delta_{ij}$ is the Kronecker delta.
Therefore, the iteration under the Newton method is given by
\begin{equation} \label{e:newton}
	\ww \;\leftarrow\; \ww + \Delta\ww \qtext{for} \Delta\ww = - \left[ \nabla \vect{F}(\ww) \right]^{-1} \vect{F}(\ww)
\end{equation}
However, unlike the CCD method, the Newton method iteration cannot guarantee that the converged weights are positive. \citet{spinu2013algorithm} overcomes the problem with the damped Newton method,
$$ \ww \;\leftarrow\; \ww + \eta\Delta\ww 
$$
where $\eta\le 1$ is a function of $\ww$ and $\Delta\ww$. Although we do not discuss the exact procedure, the basic idea is to use $\eta < 1$ in the early stage when $\ww$ is away from the solution to ensure $w_i>0$ and $\eta = 1$ later when $\ww$ is close enough to the solution for faster convergence. \citet{spinu2013algorithm} use Eq.~\eqref{e:cond_ccd} for the initial estimation of the Newton method.

Our enhancement of the Newton method is based on the initial estimation. Using an initial guess closer to the solution, we aim to use $\eta=1$ throughout the iteration, without converging to a negative weight. If this can be achieved, one can use generic Newton method routines available in many numerical analysis packages that are highly optimized for the system.

We improve the original initial guess, Eq.~\eqref{e:cond_ccd}, by updating it through the one-step CCD iteration, Eq.~\eqref{e:ccd2}. However, instead of a slow cyclical update, we use the \textit{batch} update, where the old $w_i$ values are used on the right-hand side:
$$ w_i = \sqrt{a_i^2 + b_i} - a_i \qtext{for}
a_i = \frac{\sum_{j\neq i} R_{ij}}{2\sqrt{\sum_{j,k} R_{jk}}}.
$$
This new initial guess is more efficiently computed in the vectorized form, 
\begin{equation} \label{e:init_newton}
	\ww = \sqrt{\vect{a} * \vect{a} + \vect{b}}\; -\; \vect{a} \qtext{for}
	\vect{a} = \frac{\left(\Corr \vect{1}_N - \vect{1}_N\right)}{2\sqrt{\vect{1}_N^T \Corr \vect{1}_N}}.
\end{equation}
The numerical experiments in the next section demonstrate that our new initial guess is effective such that the method converges to a positive weight for almost all randomly generated test cases.

\section{Numerical experiment} \label{s:numeric} \noindent
We tested the numerical performance of the improved algorithms.\footnote{The tests are performed in Python on a computer running the Windows 10 operating system with an Intel Core i5--6500 (3.2 GHz) CPU.} For comparison, we implemented the following three algorithms:
\begin{itemize}
	\item The original CCD method, Eq~\eqref{e:ccd}
	\item The improved CCD method, Algorithm~\ref{a:ccd}
	\item The Newton method, Eq~\eqref{e:newton}, with the improved initial guess, Eq~\eqref{e:init_newton}
\end{itemize}
All methods were implemented in Python. For the Newton method, we used the generic root solver \texttt{scipy.optimize.root} function\footnote{See \url{https://docs.scipy.org/doc/scipy/reference/generated/scipy.optimize.root.html}. We used \texttt{method=`hybr'} (default) option.} in the Python SciPy package.
For all methods, we consistently use the error tolerance $\varepsilon = 10^{-6}$.

In the test, we solved the risk parity portfolio for the correlation matrices randomly generated with the \texttt{scipy.stats.random\_correlation} class\footnote{See \url{https://docs.scipy.org/doc/scipy/reference/generated/scipy.stats.random_correlation.html}.} using the Python SciPy package. The routine takes nonnegative eigenvalues as inputs. Subsequently, it uses the algorithm of \citet{davies2000numerically} to generate a random correlation matrix. We used two methods to generate eigenvalues to test both positive definite and positive semi-definite correlation matrices:
\begin{itemize}
	\item \textbf{Test 1}: all eigenvalues sampled from independent uniform random variables between 0 and 1. 
	\item \textbf{Test 2}: 80\% of eigenvalues sampled from independent uniform random variables between 0 and 1, and 20\% set to zero.
\end{itemize}
While prior studies~\citep{griveau2013fast,bai2016least} typically test only positive definite case (i.e., \textbf{Test 1}), we think that the positive semi-definite case (i.e., \textbf{Test 2}) is also important because it is often encountered in practice when the covariance is estimated from a time series. When the covariance between $N$ assets is estimated from $M$ time periods with $M<N$, the estimated covariance matrix has at most $M$ strictly positive eigenvalues.

\begin{figure}[tbp]
	\caption{The computation time in linear (top) and log-log scales (middle), and the number of iterations (bottom) for randomly generated positive definite correlation matrices (\textbf{Test 1}). The values are the averages of 200 tests for each $N$.
	\label{f:time-pos}}
	\begin{center}
		\includegraphics[width=0.6\linewidth]{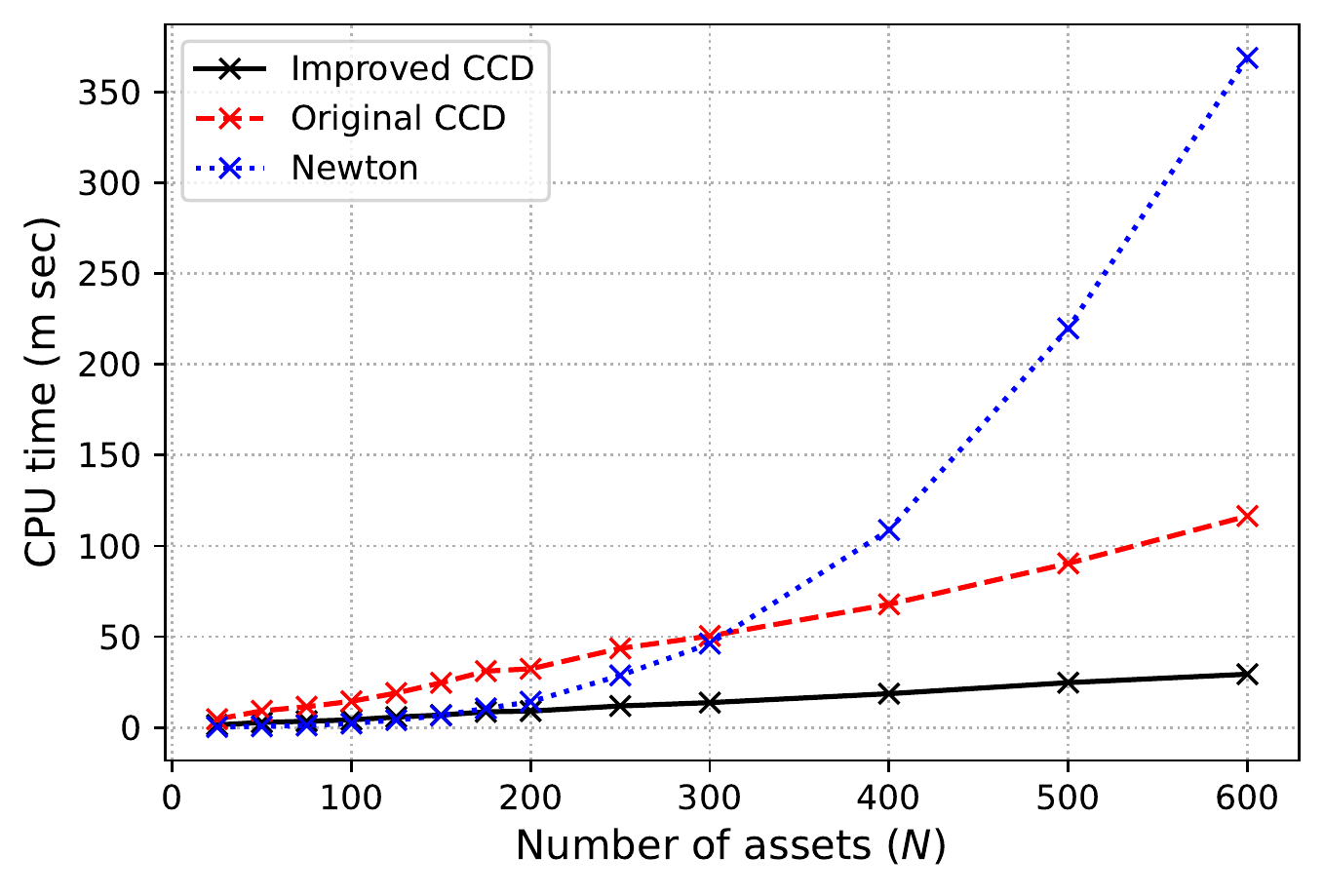}\\
		\includegraphics[width=0.6\linewidth]{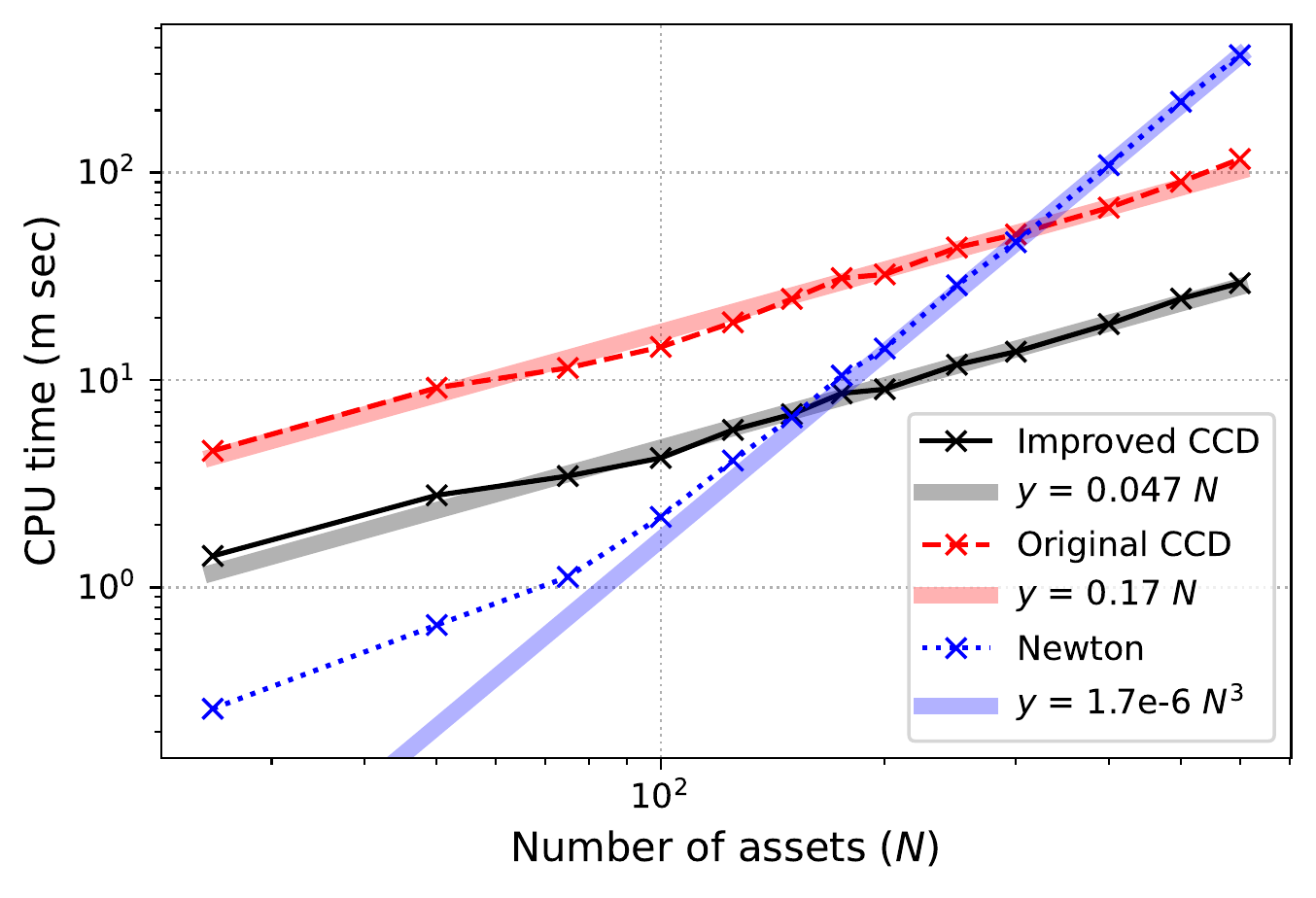}\\
		\includegraphics[width=0.6\linewidth]{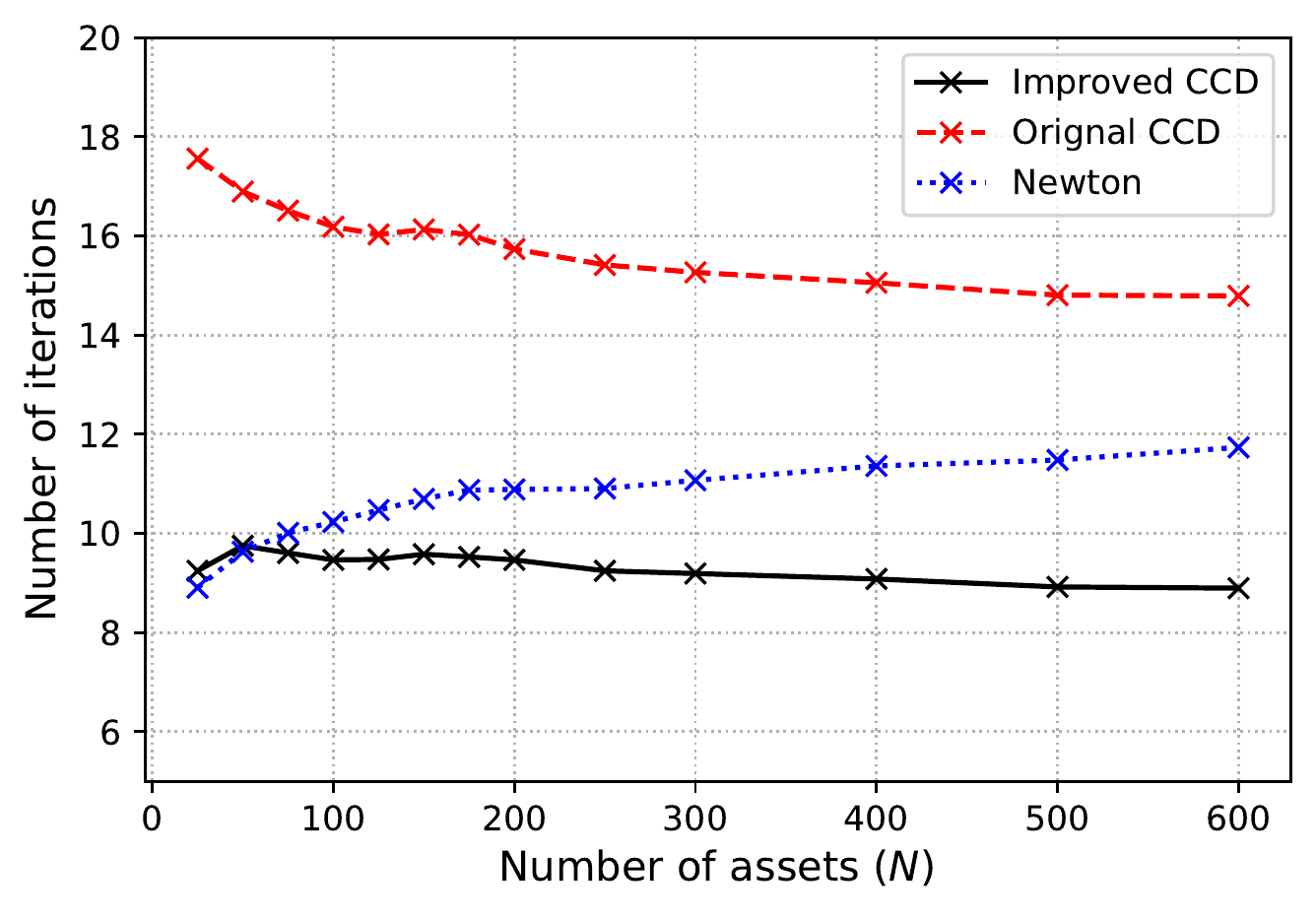}
	\end{center}
\end{figure}

\begin{figure}[tbp]
	\caption{The computation time in linear (top) and log-log scales (middle), and the number of iterations (bottom) for randomly generated positive semi-definite correlation matrices (\textbf{Test 2}). The values are the averages of 200 tests for each $N$. \label{f:time-semi}}
	\begin{center}
		\includegraphics[width=0.6\linewidth]{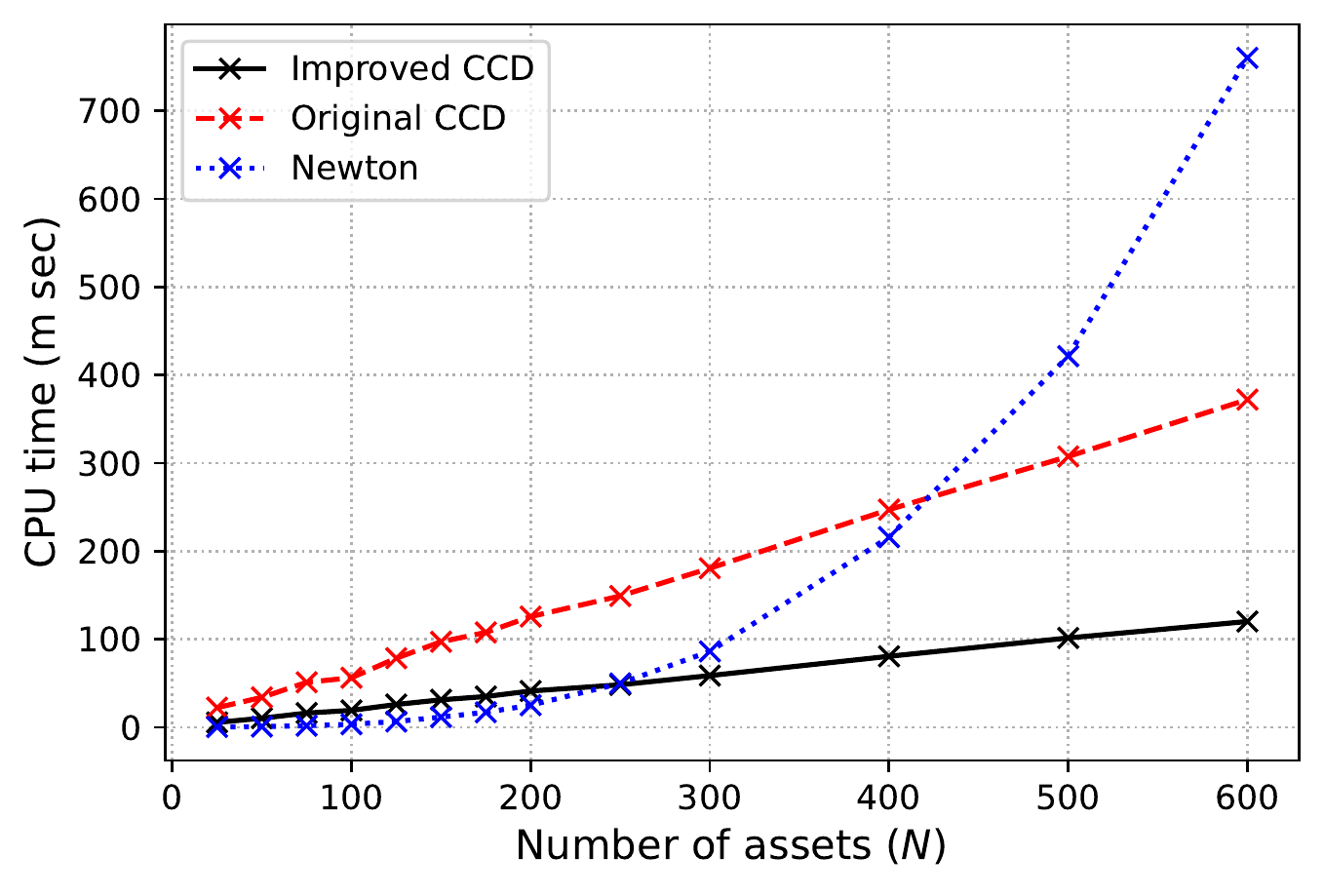}\\
		\includegraphics[width=0.6\linewidth]{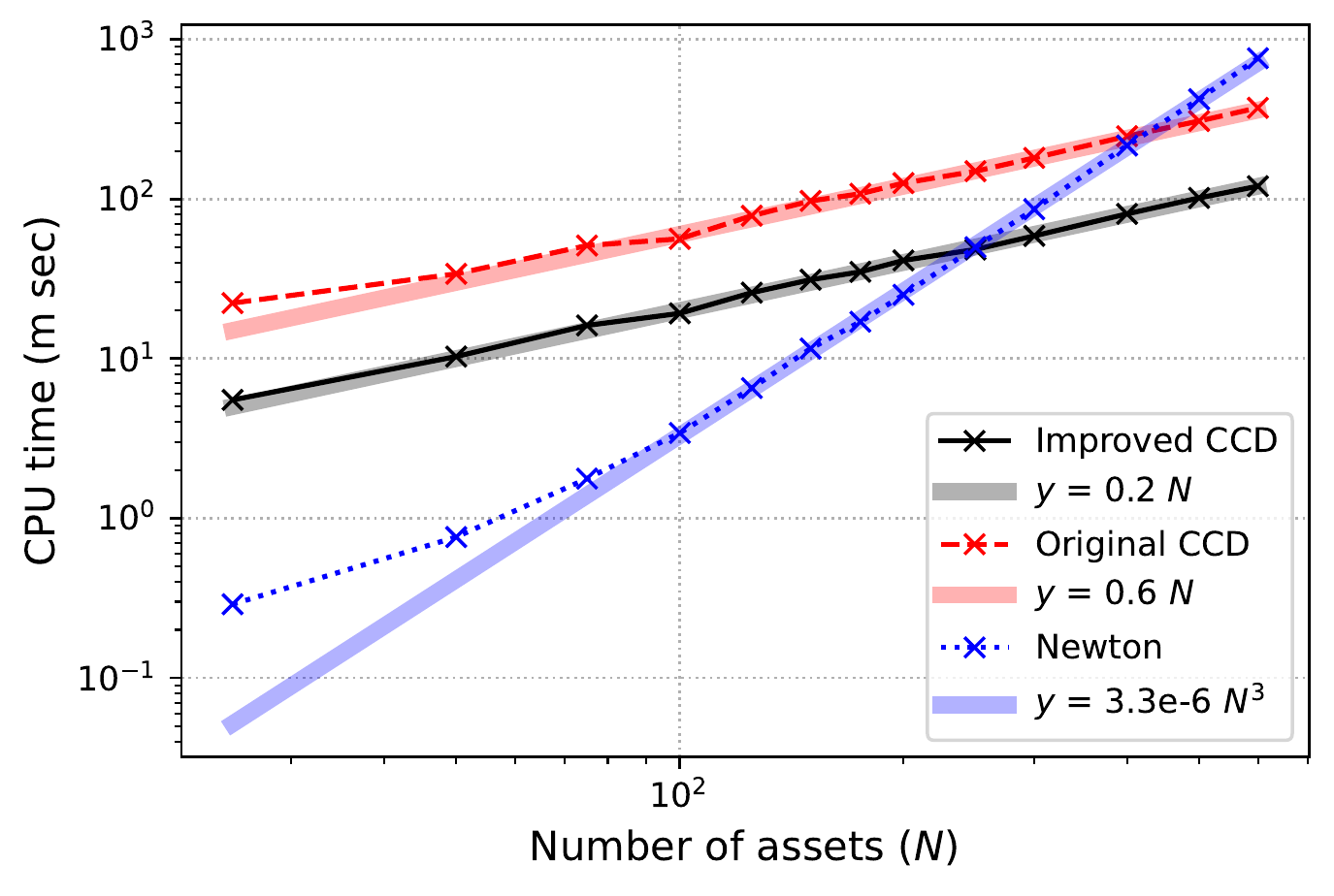}\\
		\includegraphics[width=0.6\linewidth]{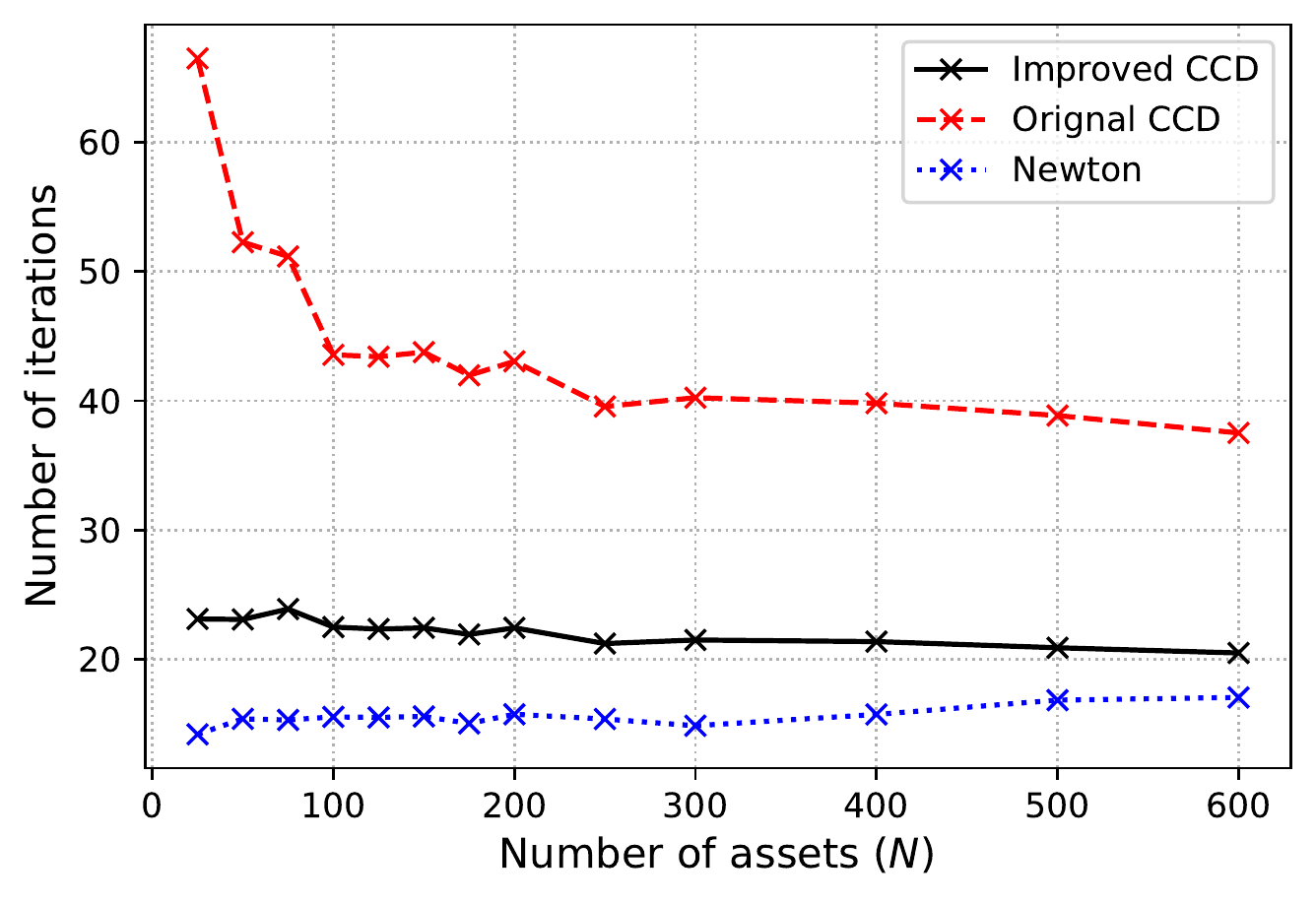}
	\end{center}
\end{figure}

Figure~\ref{f:time-pos} shows the computation time and number of iterations for \textbf{Test 1}. Several observations are in order. First, our improved CCD method performs the best in terms of both CPU time and iterations. Although the Newton method is marginally faster than the improved CCD method for $N\le 150$, the required computation time is very short anyway for those small $N$. 

Second, our improved CCD method is three to four times faster than the original CCD, saving approximately 6.5 iterations on average. Although not reported in the figure, we also ran the improved CCD method without the rescaling step in Eq.~\eqref{e:ccd2-rescale} to measure the gain from rescaling. The rescaling step saves about 1.5 iterations on average.

Third, the Newton method successfully converges to positive weights in all cases. This confirms that the damped Newton method is unnecessary with an improved initial guess, Eq.~\eqref{e:init_newton}. Nevertheless, the Newton method is inferior to the improved CCD method.
The log-log plot clearly shows that the computation time scales as $O(N^3)$ for the Newton method, but scales as $O(N)$ for both CCD methods. The $O(N^3)$ scale of the Newton method appears to be related to the computation of the Jacobian inversion, $\left[ \nabla \vect{F}(\ww) \right]^{-1}$. In the optimized multidimensional Newton method, inversion is not computed in every iteration. In our test, inversion is typically computed only once, that is, at the initial guess. However, Jacobian inversion still slows the Newton method for a large $N$. 

Figure~\ref{f:time-semi} shows the results for \textbf{Test 2}. While the relative performance between the three methods is similar, the overall computation becomes slower than \textbf{Test 1}, requiring more iterations. This indicates the difficulty of solving the risk parity portfolio against the positive semi-definite covariance matrices. Moreover, the Newton method shows instability. It fails in convergence for two cases and converges to negative weights for 13 cases. Conversely, the CCD methods stably converge to positive weights for all cases. 

We believe from the numerical tests that our improved CCD method is a fast and stable method for solving risk parity weights.

\section{Conclusion} \label{s:con}
With the growing popularity of the risk parity model, several numerical methods have been proposed to solve portfolio allocation. We present improvements to two existing methods based on iterations: the CCD and Newton methods. Numerical experiments show that the improved CCD method performs the best in terms of speed and stability.

\newpage
\singlespacing
\bibliography{RiskParity}
\end{document}